# Open Problems in Human Trait Genetics

Nadav Brandes[1], Omer Weissbrod[2], Michal Linial[3]

[1]School of Computer Science and Engineering, The Hebrew University of Jerusalem, Jerusalem, Israel; [2]Department of Epidemiology, Harvard T.H. Chan School of Public Health, Boston, MA, USA; [3]Department of Biological Chemistry, The Alexander Silberman Institute of Life Sciences, The Hebrew University of Jerusalem, Jerusalem, Israel.

## Abstract

Genetic studies of human traits have revolutionized our understanding of the variation between individuals, and opened the door for numerous breakthroughs in biology, medicine and other scientific fields. And yet, the ultimate promise of this area of research is still not fully realized. In this review, we highlight the major open problems that need to be solved to improve our understanding of the genetic variation underlying human traits, and by discussing these challenges provide a primer to the field. Our focus is on concrete analytical problems, both conceptual and technical in nature. We cover general issues in genetic studies such as population structure, epistasis and gene-environment interactions, data-related issues such as ethnic diversity and rare genetic variants, and specific challenges related to heritability estimates, genetic association studies and polygenic risk scores. We emphasize the interconnectedness of these open problems and suggest promising avenues to address them.

# Preface

Since the days of Gregor Mendel and his series of pea plant experiments in the middle of the 19th century, a key motivation of genetic research, perhaps the leading motivation, has been to understand the link between genotype and phenotype. When the human genome project was declared complete in 2003, hopes were high for a full understanding of human genetics and its effects on human traits. The idea to genotype large cohorts of individuals and, through simple statistical tests, compiling an atlas mapping between genes and human traits – be it diabetes, schizophrenia or height – had been around for quite some time [1]. But excitement rapidly shifted to disappointment, as efforts to find the genetic variation underlying complex human diseases ended up explaining only a small fraction of the phenotypic variance [2].

Almost two decades later, with millions of individuals genotyped across thousands of genome-wide association studies, it is now well acknowledged that things are not that simple. But it is worth asking, why are they not, actually? Why haven't we mapped most of the genetic variation underlying human traits, and why are we still unable to make accurate individual phenotypic predictions from genetic data? What are the concrete problems we are now facing, and what bottlenecks are slowing us down and preventing genetic research from unlocking its full potential? Asking these questions and attempting to answer them will allow us to make more effective progress and eventually achieve the field's long-term potential.

# Scope

The domain of knowledge we are dealing with, mapping and understanding the genetic variation underlying human phenotypic variation, is a huge area of scientific inquiry. Among its main subdomains are: (i) genetic association studies, which seek robust causal links between genetic variants and human traits [2–4], (ii) polygenic risk scores, which aim to predict traits from genetics [5], and (iii) heritability estimates, which estimate the fraction of a trait's variance that is due to genetic variation [6].

The primary applications of these research activities are twofold: (i) obtaining insight into the biological mechanism at the molecular and cellular level underlying the disease or trait under study, and (ii) making informed predictions that can be clinically useful, even in the absence of knowledge about mechanism or causality. Even if we do not understand why a certain individual is at high risk for heart disease, it is still useful to know that they are at such risk, especially if we can reliably quantify it [7].

This review only deals with topics directly related to the interface between genetics and human traits. We do not address topics related to genetics but not to phenotypic variation in present-day humans (such as functional elements in the human genome, or questions of evolution, conservation and fitness). We also do not discuss methods that use genetics only as an instrument to study relationships between traits, such as Mendelian Randomization [8], as well as technical aspects of genotyping and sequencing (including variant calling and quality control) and cancer genomics (i.e., the study of somatic mutations in tumors). Also outside of our main focus are gene expression, epigenetics and other omics.

It is common to distinguish between Mendelian and complex traits. Mendelian traits are traits that follow Mendel's laws of inheritance, in particular the first law dictating pure dominant or recessive inheritance with high penetrance. A Mendelian trait is typically linked to a single gene (sometimes a handful of genes) and it tends to be deterministically determined by genetics alone. Non-Mendelian traits, known as complex traits, show the opposite characteristics: they are typically influenced by variants across numerous (sometimes thousands) of genomic loci, each with limited effect, and they typically exhibit substantial environmental effects and tend to be nondeterministic. While we generally understand the genetics of Mendelian traits [9], the study of complex traits still poses many challenges.

This review primarily focuses on analytical challenges in studying the genetics of complex human traits. Because we are interested in methodology, we mention specific traits only as examples and illustrations of general trends, and while we focus on human, it is useful at times to gain insight from the study of other organisms.

# A list of open problems

The notion and tradition of publishing "open problems" is borrowed from the mathematical disciplines, where the promotion and explicit discussion of major open challenges has a great role in prioritizing work, sub-branching and otherwise advancing these fields. This review is an attempt to map concrete and solvable open problems related to the genetics of human traits.

Altogether, we have identified 16 major open problems which we consider important to the field (Table 1; a corresponding list with detailed explanation and selected references about each of the open problems is available in Supplementary Table S1). Since these open problems are highly interconnected, we go through them in the remainder of this review by thematic topics (rather than by the categorical ordering in Table 1).

# Genetic association studies and their limitations

Genome-Wide Association Study (GWAS) [2–4] is the most common type of genetic study. The idea behind GWAS is straightforward: independently examine each genotyped variant in the genome, and test it for statistical association with the studied phenotype. To minimize confounding by non-genetic factors (such as sex and age), statistical testing is typically done with variations of linear or logistic regression (depending on whether the studied trait is continuous or binary, respectively). The main strengths of GWAS are its simplicity and generality. Since the method makes almost no assumptions about the nature of associations and their biological basis, GWAS can in principle identify any genetic association, provided a sufficiently large cohort. The required sample size is determined by the allele frequency and effect size of each association. There are, however, two main issues limiting one from inferring a causal link in the face of a significant GWAS association: population structure and linkage disequilibrium.

## Population structure

The phrase "correlation does not imply causation" needs no rehearsing. In the case of genetics, traits only rarely affect genetics (e.g., through assortative mating [10]). This means that if we seek to draw a causal genotype-to-phenotype link based on a statistical association, our primary concern is confounding, namely the existence of other variables affecting both an individual's genetics and the trait (there is also the less trivial concern of collider bias, which is addressed later). Specifically, we know that an individual's genetics is determined by the genetics of their parents. It follows that the precise family and ethnic identity of an individual (generally simply referred to as their "population" or "ancestry") is a suspect confounder for any genetic association, as it affects the environment which an individual is born into (e.g., weather, nutrition, etc.) and, as a result, can affect the studied trait (**Fig. 1A**) [11].

To avoid spurious associations, genetic studies must account for population structure. The most common method for that is principal component analysis (PCA) [12]. It was shown that accounting for the 5-40 top principal components of genetic variation in a cohort (by including them as additional covariates in the regression analysis) can eliminate spurious results due to population structure. It is also a common practice to split cohorts by self-reported ethnic identities [13]. Another class of commonly used methods are linear mixed models, which can control for even more nuanced population structure [14, 15]. While it is crucial to account for population structure, overcorrection could lead to reduction of statistical and predictive power, and to underestimation of the role of genetics in traits. Focusing on homogenous ethnic groups also underplays the potential genetic basis of phenotypic differences between groups.

Despite the capacity of existing methods to account for most of the phenotypic variance that is due to population structure, there are increasing concerns that residual population structure can still lead to spurious associations, especially with very large cohorts that are now emerging (**open problem #1: population structure**; **Fig. 1B**) [16]. For example, unaccounted population structure led researchers to mistakenly conclude that height-associated genetic variants were under selection in the European population [17]. It is still not fully understood how residual population structure affects the results of GWAS and polygenic risk scores. More research is also needed for establishing new methods and practices to address the problem. Two strategies seem particularly promising: (i) studying more diverse ethnic populations (especially non-Europeans), and (ii) using family-based designs for genetic studies.

The absence of ethnic diversity in genetic studies is hard to overemphasize (**open problem #7: ethnic diversity**). Individuals of European ancestry are overwhelmingly overrepresented in contemporary cohorts and, as a result, most published results do not directly transfer to other ethnic groups [18]. The immediate implication is that many of the benefits of genetic studies are of little benefit to most of the world's population. Moreover, studies that are constrained to only one ethnicity may suffer from residual population structure driven by subpopulations of that ethnic group (e.g., across different geographic regions in Europe) [16]. Hence, ethnic diversity is also an incredibly powerful strategy for avoiding spurious associations driven by population structure [13]. The reason is that, while it is entirely possible for a variant to be spuriously correlated with a trait within one ethnic group due to specific subpopulations, it is unlikely to observe the same trend in an entirely different ethnic group. Thus, the replication of a genetic association across multiple ethnicities is strong evidence for causality. In addition to more diverse datasets and increased awareness by researchers, there is also need for better analytical methods for robust analysis of trans-ethnic cohorts [13].

Another promising strategy to account for population structure is family-based studies [19]. As an illustration, consider the case of trios, where the two parents of each studied individual are also genotyped. When testing an autosomal variant, one can focus on trios where at least one of the parents is heterozygous for the tested variant. According to Mendel's laws of heredity, the child is then expected to have exactly 50% chance of inheriting the variant from that parent and, crucially, this happens completely at random. By testing the fraction of transmitted variants among cases and controls (or, in the case of a continuous phenotype, the correlation between the phenotype and the fraction of transmitted variants), one can establish a robust association. Since Mendelian transmission is random, once the parents' genetics is accounted for, this study design avoids the problem of population structure altogether. Like trios, more complex family structures can also be utilized in similar ways [19]. On the downside, families are harder to recruit than unrelated individuals. In addition, each trio comprises only a single datapoint, meaning that the effective sample size is only a third of the number of genotyped individuals.

While these days family-based studies are mostly used in the context of Mendelian traits, we argue that they should play a greater role in the study of complex traits [20, 21]. It is becoming increasingly evident that certain questions in human genetics can be decisively answered only through this study design (Supplementary Table S1). Accessible resources with an abundance of related individuals, on the scale of contemporary large-scale biobanks [22, 23], with rich genotypic and phenotypic data, could be a major force advancing the field. Unfortunately, such resources are still scarce (**open problem #6: family-based vs. population-based cohorts**). Contemporary family cohorts are not easily accessible, and are typically restricted to specific phenotypes (e.g., [24]). Also, methods combining and extracting signal from both family- and population-based data could be highly beneficial to the field.

**Table 1: Open problems**

| Category | # | Open problem |
|---|---|---|
| General | 1 | Population structure |
| | 2 | Non-additive and epistatic genetic effects (GxG) |
| | 3 | Gene-environment interactions (GxE) |
| Data | 4 | Rare variants |
| | 5 | Non-standard genetic variation |
| | 6 | Family-based vs. population-based cohorts |
| | 7 | Ethnic diversity |
| | 8 | Phenotype definition |
| | 9 | Selection bias |
| Heritability | 10 | Heritability estimate interpretation |
| | 11 | Missing heritability |
| Association studies | 12 | From association to causality |
| | 13 | From causality to mechanism |
| Polygenic risk scores | 14 | Genotype-to-phenotype prediction performance |
| | 15 | The clinical utility of polygenic risk scores |
| | 16 | Model transferability |

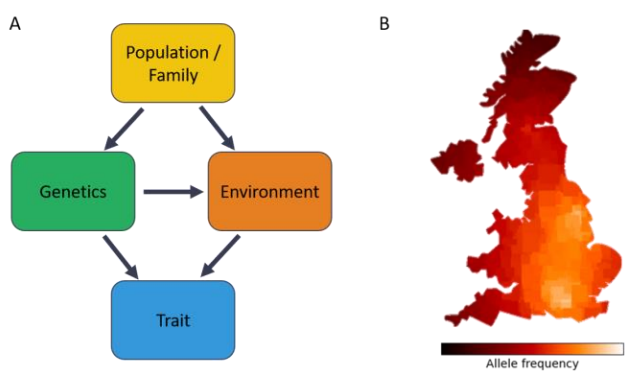

**Fig. 1: Population structure confounds human genetic studies**

(**A**) The population that an individual is born into influences their genetics and their environment, which are the two components affecting traits. As a result, genetic associations with human traits are confounded by population structure. (**B**) Even when considering a specific ethnic group and controlling for the major axes of genetic variation in a cohort, the allele frequency of some variants can still vary across populations and exhibit clear geographic patterns, a problem known as "residual population structure".

## Linkage disequilibrium and the pursuit of causal variants

Linkage disequilibrium (LD) is the phenomenon that, due to inheritance through homologous recombination, individuals with a specific allele of one variant are likely to have a specific allele of a neighboring variant. A combination of multiple alleles that appear together in an individual across neighboring variants in a genomic region is often referred to as a haplotype. The resulting LD structure of correlated variants and haplotypes has profound implications for genetic studies [3].

In some ways, the phenomenon of LD is quite convenient. LD allows genotype imputation, namely the deduction of an individual's genotype across a large set of genetic variants from a much smaller set of variants. Genotype imputation relies on haplotype reference panels which provide the full haplotypes of a large cohort of individuals (e.g., 32,488 individuals in the Haplotype Reference Consortium [25]). Like most genetic resources, haplotype reference panels are highly biased towards individuals of European ancestry, and so are the popular DNA microarrays (which usually genotype variants present in these populations). As a result, the genetic coverage obtained for non-Europeans is typically much lower (open problem #7: ethnic diversity) [13]. Additionally, the common use of external haplotype reference panels to account for LD in published GWAS results (e.g., to conduct a meta-analysis) yields biased estimates of the actual LD patterns in the original dataset, and negatively affects downstream analysis [26]. Publishing the full pairwise LD matrices together with the rest of the GWAS summary statistics could be a beneficial norm [27].

Due to LD, genetic variation has far fewer degrees of freedom than would naively be assumed by the total number of variants in the human genome. Because of that, adjusting the p-values that are independently derived for each tested variant in GWAS using Bonferroni correction would be too conservative. Instead, it is assumed that the LD structure in human is such that an individual's genotype has roughly 1M degrees of freedom, meaning a genome-wide significance threshold of 5E-08 (or an exome-wide significance threshold of 5E-07, derived from roughly 100K degrees of freedom). While these thresholds are the norm in genetic studies, more stringent cutoffs have been recommended under certain conditions [28].

A negative consequence of LD is its obscuring of causal variants. Even when a genomic locus is robustly established as causally linked to a phenotype, it is very difficult to tell with certainty which of the variants in that locus are behind that causal link, a task known as fine-mapping (**Fig. 2A**). Numerous fine-mapping methods have been developed [26, 29–31]. These methods often appeal to Bayesian reasoning and rely on functional genomic annotations, under the assumption that causal variants are more likely to be located in functional sites of the genome. Despite these efforts and the great progress made, it is still an open problem how to establish the causality of a variant or a gene with certainty (**open problem #12: from association to causality**). In addition to method development, there is much room for the formalization of standards for establishing causality for research or clinical purposes [32].

Trans-ethnic genetic studies can be very useful for fine-mapping, in a similar way to their utility in dealing with population structure (open problem #7: ethnic diversity). Since different populations show different LD structures, a trans-ethnic meta-analysis will generally point at a much smaller set of candidate causal variants within a genomic region (**Fig. 2B**) [4, 13, 29]. However, if a genetic effect is unique to a specific population (e.g., due to interactions with other genetic or environmental factors), trans-ethnic analysis will not fine-map it effectively.

Another useful approach to pinning down the causal elements underlying genetic associations is to abandon the pursuit of specific variants and shift our attention to other elements of the genome such as genes and pathways. Gene-based methods have become increasingly popular in recent years [33–37]. For example, there are methods detecting genes that are affected by variants influencing their expression to different levels in diseased cases compared to healthy controls [35, 36]. Among the merits of gene-based methods are reduced burden of multiple testing, easier interpretation of associations, and the ability to aggregate signal spread across many variants. If distinct variants that are not in LD perturb the same gene (e.g., change its coding sequence or expression level) and are associated with the same trait, then this could comprise an overall stronger evidence for the causality of that gene. However, gene-based approaches are still susceptible to LD; it is possible that signal from nearby variants would leak into variants affecting the gene. Gene-based methods are also sensitive to modeling assumptions and the specific details of how they aggregate the signal spread across variants.

An important factor limiting our progress towards established standards for causality is the shortage of confirmed, experimentally-validated causal variants that can be used as a gold standard for validation and evaluation of methods. As a result, evaluation of methods in the field is typically based on computational simulations (which are sensitive to modeling assumptions). In addition to empirical benchmarks, the research of causal variants could benefit from well-designed competitions (similar to the competitions held by the protein research community [38, 39]).

Genetic studies have provided scientists across a range of disciplines, including the social sciences, with powerful tools to answer causal questions that interest them (e.g., through Mendelian randomization [8]). However, it is important to be aware that these tools can still be biased.

## Direct vs. indirect effects

Even when a variant or a gene is proven causal, its phenotypic effect might be indirect. Direct genetic effects influence traits within the individuals who carry the causal alleles, while indirect genetic effects influence their relatives. Important types of indirect genetic effects include (i) parental effects (that influence traits in children through their parents), (ii) assortative mating (i.e., influences on mate choices, which in turn can lead to further parental effects), and (iii) sibling effects (which may also influence traits through the environment). While both direct and indirect genetic effects are causal, the distinction between them is often crucial. For example, when seeking drug targets to treat afflicted individuals, only direct genetic effects are expected to be relevant. Common analytical methods such as GWAS do not distinguish between direct and indirect genetic effects, and it is still very much an open challenge to effectively dissect the two. Family-based studies that control for parental genotypes, on the other hand, in addition to being useful for separating causal from confounding effects, can be used to further separate causal effects into direct and indirect effects (open problem #6: family-based vs. population-based cohorts) [40].

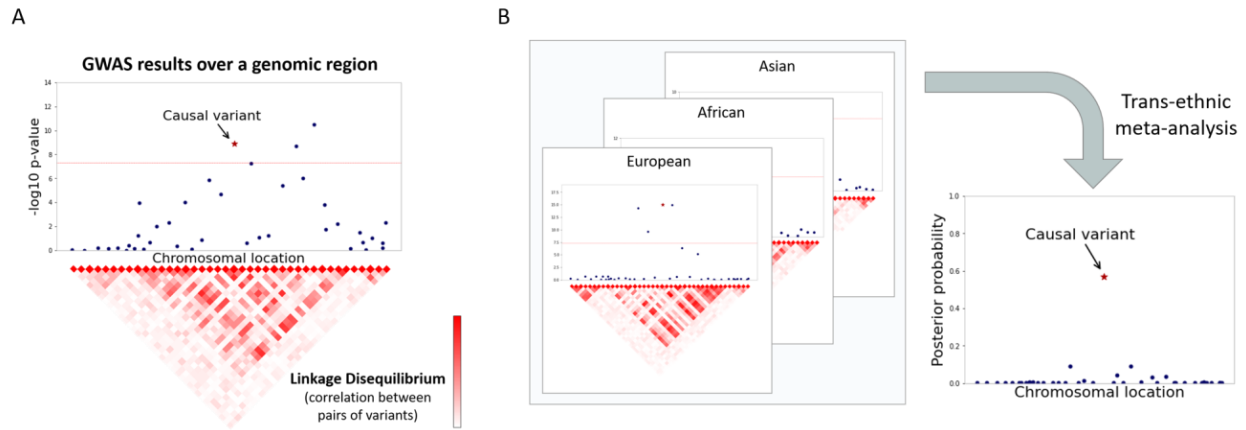

**Fig. 2: Identifying causal variants in the presence of linkage disequilibrium**

(**A**) A single causal variant is in linkage disequilibrium with other nearby variants. As a result, variants that are correlated with the causal variant also obtain significant p-values even though they are not causal. (**B**) Combining GWAS summary statistics from three different ethnic groups, each exhibiting a different linkage-disequilibrium pattern, to fine-map the results. By assuming that only one of the variants is causal, it can be recovered with high confidence.

## From causality to mechanism

Even among genetic effects for which causality is established, only a small fraction have a known, well-established biological mechanism (**open problem #13: from causality to mechanism**) [41, 42]. As most causal variants are in non-coding regions and exert weak effects, it is challenging to understand how they affect traits [4]. Without understanding how a variant affects a disease, it is of little use in providing biological insight on its etiology and progression.

At the beginning of the GWAS era researchers were struggling to identify robust associations. It appears that nowadays we have the opposite problem: we are flooded with hundreds of thousands of genetic associations that we don't really know what to make of [4]. It was actually suggested that many complex traits are not only polygenic (i.e., genetically driven by many genes), but could in fact be considered omnigenic, namely that they are affected by most of the genes in the genome [43]. For example, it was estimated that more than 71% of the 1 Mbp regions in the human genome contain at least one schizophrenia risk variant [44]. In such cases, additional mapping of genomic loci associated with omnigenic traits, without considering other factors (such as effect sizes), is not expected to substantially contribute to our understanding of mechanism.

One of the limitations in establishing causality and discerning mechanism that is specific to human genetics is our inability to run controlled experiments, which are an indispensable tool in plants and animals. There are however several strategies to utilize experimental methods in human genetic studies. One approach is to validate human genetic associations in other model organisms (e.g., by knocking out a homologous gene and observing the phenotypic effects on the animal). Another strategy is to look at cellular and tissue-related phenotypes for which experiments can be conducted (e.g., using primary culture, human cell lines or induced stem cells) as a proxy for the studied phenotype. We now possess high-throughput experimental methods that can not only functionally annotate fixed features of the genome (such as introns and promoters), but also detect the variable effects of genetic variation. For example, Perturb-seq is a recently developed method combining CRISPR-based genetic perturbations with single-cell RNA sequencing to detect variants that causally affect gene expression [45]. Using such experimental methods, we can observe the effects of genetic variants on cells and postulate on the mechanism of their effects on human traits. Unfortunately, experimental methods are still dramatically more expensive (in time, labor and money) than analytical methods. They also force additional decisions such as which cell types to study. Moreover, as many complex diseases and conditions are manifested at the organism level, without known cellular-level indicators, cell-based methods could be inadequate. In light of these challenges, devising experimental and analytical strategies for finding or guiding the search for biological mechanism would be of tremendous benefit.

# Polygenic risk scores: unfulfilled potential

Considering the limitations of GWAS and the fact that even variants that are proven causal tend to exert only weak effects, raw GWAS results provide limited utility for predicting complex traits (e.g., to detect individuals who are at high risk of developing type 2 diabetes). To make a meaningful prediction about a complex trait, one has to aggregate the weak signals spread across many genomic regions. An analytical model aggregating an individual's genotype into an overall phenotypic prediction is called a polygenic risk score (PRS) [5]. In essence, this can be seen as a machine-learning task: training a model that takes an input x (one's genotype, and optionally other variables) and outputs a prediction y (the individual's predicted trait value). It should be noted that some define PRS as strictly linear models (which aggregate variants by multiplying them with learned weights and summing the multiplication terms), but we prefer to define PRS more generally as any genotype-to-phenotype predictive model.

PRS can be trained in various ways. Given a sufficiently large cohort of genotyped and phenotyped individuals, one can train a PRS from scratch using standard machine-learning algorithms on individual-level data. More commonly, summary statistics of published GWAS results are meta-analyzed into a linear model.

Accurate and reliable PRS have the potential to transform healthcare. Many common diseases (including metabolic, psychiatric, autoimmune and cardiovascular diseases) have a very substantial genetic component [46]. Knowing that individuals are at elevated or reduced risk for a specific disease could prioritize screening and follow-ups, guide diagnosis and inform medical interventions [7]. However, despite over a decade of refining models with exponentially increasing cohort sizes, the predictive power of most PRS is still quite poor (**open problem #14: genotype-to-phenotype prediction performance**). While there has been some success in genetic prediction of specific phenotypes (such as height [47]), most diseases and clinically relevant traits are still far from the full potential of genetics-based risk assessment; the phenotypic variance explained by existing models is only a small fraction of the trait heritability, and most models do not reach clinical relevance [7]. Many of the potential reasons for the unsatisfying performance of PRS are linked to the question of missing heritability and to questions surrounding nonlinear genetic effects, which are addressed later. It remains an open question whether better methodology would allow us to substantially improve PRS performance without larger cohorts. A useful practical approach to improving PRS performance is to incorporate clinical factors into the predictive models on top of genetic markers (e.g., using body mass index and birth weight to improve type-2-diabetes PRS [48]).

Another problem of PRS is in their capacity to generalize from the cohort they have been trained on to other settings, including different populations and genotyping technologies (**open problem #16: model transferability**). This is one of the main barriers for deploying these models in real-world clinical settings. In PRS it is usually not necessary to pin down the causal variants (as non-causal variants that are only in LD with the causal variants can provide the same predictive power), but this may negatively affect model transferability due to population-specific LD patterns [18, 49]. Other dissimilarities between populations that can limit model generalization include differences in allele frequencies or effect sizes, or different interactions of the genetic effects with the environment or other genetic factors. Like with many other problems in genetic research, here too it is extraordinarily helpful to use trans-ethnic data when training or fine-tuning PRS (open problem #7: ethnic diversity) [13]. When a PRS is transferred from one dataset to another, even when both are in principle of the same population, substantial declines in performance are still common [50]. The instability of PRS is often attributed to batch effects and to differences in genotyping technologies across datasets, but we still lack a good understanding of the reasons for these sensitivities. At present, we do not have a good theoretical framework estimating how much accuracy loss we should expect when transferring PRS between settings. Even more importantly, there is a great need of methods and strategies that would make PRS more robust and reliable. Such strategies could include adjustment and calibration of models to new settings, or training them to be more robust in the first place. Perhaps the ongoing revolution of causal inference could play a role in training PRS that capture causal signals instead of merely statistical associations which, in addition to many other benefits, would be more robust [51].

# Genetics in the clinics: are we there yet?

Genetic tests are routinely used in many clinical settings, including parental screening, diagnosis of children with developmental disorders and drug prescription [32, 52]. To diagnose Mendelian traits, genetic counselors look for pathogenic variants that explain the observed phenotype and clinical history of the patient. However, a known variant with affirmed pathogenicity is not always found. For example, the Mendelian disease could be the result of a rare or de-novo variant not previously reported. In the absence of conclusive pathogenic variants, genetic counselors may resort to more circumstantial evidence, such as the predicted functional effects of variants based on computational algorithms. The American College of Medical Genetics and Genomics recommends a five-tier system of classification for variants relevant to Mendelian disease, based on the strength of evidence supporting their role in the disease: (1) pathogenic, (2) likely pathogenic, (3) uncertain significance, (4) likely benign, or (5) benign [32].

Another potential use of genetics is in drug prescription. Pharmacogenetics (sometimes referred to as pharmacogenomics) studies genetic variation underlying individual response to medication, which could be used to predict (i) individual drug dose, (ii) absence of response to a drug, or (iii) individuals at serious risk of toxicity if a drug is prescribed. For example, it is now a routine clinical protocol to test for the HLA-B*57:01 allele before prescribing the anti-HIV drug abacavir, which could lead to hypersensitivity reactions in carriers of this allele. However, current uses of pharmacogenetics as part of standard medical care are still limited to a small number of well-established associations [52].

In contrast to its immense utility for Mendelian traits, genetics is rarely used as part of routine clinical practice when dealing with complex traits. This is an enormous unfulfilled opportunity, since complex diseases comprise most of the burden of diseases in developed countries, and there is a lot to gain from early intervention. The GWAS literature reporting on over 200,000 genotype-phenotype associations is almost never directly used by clinicians. Clinical practice calls for a much stronger burden of proof, and the weak effect sizes associated with most GWAS results do not generally justify substantial deviations from routine medical care. Some exceptions exist, mostly related to cancer predisposing (e.g., testing variants in BRCA1 and BRCA2 to screen for breast and ovarian cancer [53]).

Unlike raw GWAS results, PRS have the potential to revolutionize healthcare. But, as of today, PRS are not yet adopted in clinical settings in a meaningful scale [7, 49]. To be accepted by the medical community, the clinical utility of PRS should first be established (**open problem #15: the clinical utility of polygenic risk scores**). Part of the problem is the aforementioned issues of PRS having, for the most part, limited predictive power (open problem #14: genotype-to-phenotype prediction performance) and generalizability (open problem #16: model transferability). On top of that, proving the reliability and robustness of PRS-guided protocols over classic medical protocols is a complex process which could require randomized-controlled trials. Expensive clinical trials are mostly required in cases where risky policy changes are attempted (for example, delaying mammographic screenings beyond the recommended age for low-risk women). When genetics is only used to take extra precautions (for example, undergoing colonoscopy screening earlier in life than would otherwise be recommended), there is usually no need for full clinical trials, but health providers would still want to see good evidence that the decision is sensible and cost-effective. When considering the clinical utility of PRS, it should be kept in mind that they need not provide clear predictions for all or even most individuals. Identifying a subset of individuals who are at elevated or reduced risk for disease is sufficient in many settings [7]. Above all, PRS should be seen as a supplement, not replacement, for traditional risk prediction models [49].

In the long run, genetic studies (and PRS in particular) could have additional clinical applications, for example in screening human embryos for complex diseases or other polygenic traits (which raises both ethical and practical concerns) [54, 55].

# Heritability estimates and controversies

## What is heritability and why is it important?

Human genetics literature is filled with estimates of heritability for various human traits based on a variety of methods. For example, a recent study on the heritability of 551 complex traits based on the UK Biobank estimated height to be around 70% heritable and fluid intelligence score around 25% in the white British population [46]. Another study estimated the heritability of schizophrenia at 80% (based on the Nationwide Danish Twin Register) [56]. But what do these numbers actually mean?

The heritability of a continuous trait is defined as the fraction of the trait's variance that is due to genetic variation [6]. Mathematically, it is defined as $H^2 = Var(G)/Var(P)$ where $Var(P)$ is the overall trait variance of the trait and $Var(G)$ is the variance of the genetic component of the trait. While this definition appears simple, it involves certain assumptions and nuances that are easy to miss or misinterpret. First, this definition does not naturally apply to binary traits. To define the heritability of a binary trait (e.g., schizophrenia), it is typically assumed that there exists a continuous liability score underlying its manifestation (i.e., the trait manifests when the liability score is above a certain threshold). The heritability of the binary trait is then defined as the heritability of that latent continuous score [57]. Another important nuance is that heritability is, by definition, context-specific and not a fundamental property of nature; it is defined for a specific population in a specific time and place. For example, as societies become wealthier human height becomes less contingent on access to nutrition and, as a result, it is expected that more of the variance would be the result of genetic differences. As a result, we expect to see human height becoming more heritable. Another crucial question about this definition (which we address later) is what exactly is meant by "the genetic component of the trait" ($G$).

There are many motivations for estimating the heritability of a trait. Primarily, it is very useful for guiding genetic research. For example, it could be used to anticipate the theoretical limit of the performance of PRS, letting us know how far we are from a complete analytical model of a trait's genetic architecture. Likewise, heritability estimates often provide justification for further genetic studies. From a clinical perspective, it can inform family members of their risk to develop a disease diagnosed in their relatives. Heritability sometimes also arises in the face of social questions. For example, there is an ongoing debate in economy to what extent income inequality and social mobility are determined by socioeconomic status at birth vs. immutable genetic factors (as well as indirect effects from the genetics of the parents) [58]. Another motivation for heritability estimates is simple curiosity: studies proclaiming to what extent different traits are genetic often attract the public's attention.

## Methods for estimating heritability

As alluded to, explicitly modeling the genetic component of a trait through PRS generally provides only a weak signal. Therefore, estimating the variance of the genetic component of a trait requires more sophisticated methods. There are many heritability estimation methods. While they often involve complex mathematical analysis, the underlying principle shared by all methods is straightforward: the more heritable a trait is, the more we expect individuals who are genetically similar to also be phenotypically similar. By making certain assumptions about the genetic architecture of the trait and analyzing the association between genetic and phenotypic similarities, a trait's heritability can be estimated. There are three main categories of heritability estimation methods: (i) twin studies, (ii) genomic relatedness and (iii) family-based methods (**Fig. 3**). Each methodological category is suited for distinct types of data, is subject to different assumptions and has unique strengths and weaknesses.

The classic method for estimating heritability is through twin studies (**Fig. 3A**) [59]. In such studies, monozygotic (identical) twins are compared to dizygotic (non-identical) twins. If a trait is heritable, we expect monozygotic twins to be more phenotypically similar than dizygotic twins. Knowing that monozygotic twins are 100% genetically similar, whereas dizygotic twins have only 50% chance of having the same parental allele transmitted to them (in variants that are heterozygous in one of the parents), Falconer's formula can be derived. The formula states that the heritability of the trait is $H^2 = 2(r_{MZ} - r_{DZ})$, where $r_{MZ}$ and $r_{DZ}$ are the phenotypic correlations between monozygotic and dizygotic twins, respectively [60]. Notably, twin studies make the assumption that monozygotic twins do not share a more similar environment than dizygotic twins. It also assumes the lack of genetic (GxG) and gene-environmental (GxE) interactions (which we discuss later). The greatest strength of this method is that, unlike all other methods, twin studies do not require genetic data at all, only phenotypic measurements. For this reason, twin studies had provided heritability estimates long before genetic sequencing became a viable technology.

A second class of methods for estimating heritability, which has become popular since the era of GWAS, leverages genetic cohorts of unrelated individuals and includes methods such as GREML (Genomic Relatedness Restricted Maximum Likelihood) and LDSC (Linkage Disequilibrium Score Regression) [61]. GREML, the first of these methods, estimates heritability by calculating and comparing all pairwise similarities between a cohort's individuals with respect to both genotype and phenotype (**Fig. 3B**). An important weakness of these methods is their sensitivity to population structure and other environmental biases [21]. As in GWAS, population structure is commonly accounted for in GREML and LDSC by using the top principal components of genetic variation. However, there are concerning indications that residual population structure could still bias their results (open problem #1: population structure) [16, 21].

A third class of methods utilizes family-based data on genotyped relatives [62]. The key idea here is to assess whether related individuals with greater than expected genetic similarity also share a greater than expected phenotypic similarity (**Fig. 3C**). By accounting for the genetics of parents, family-based methods are insensitive to most environmental biases undermining twin and cohort studies, including population structure. The family-based approach also captures only direct genetic effects (unlike cohort-based methods which also capture indirect effects) [40]. On the downside, it is harder to obtain a sufficient number of samples required for accurate heritability estimates.

## Broad- and narrow-sense heritability

A crucial difference between heritability estimation methods is the type of heritability they measure. Earlier we gave the definition of $H^2$, known as the broad-sense heritability. This is the type of heritability that twin studies presume to measure (although, as we later discuss, genetic interactions and environmental biases can lead to overestimation of the true heritability). On the other hand, methods based on genotype data typically measure something slightly different, known as the narrow-sense heritability (or SNP heritability), denoted by $h^2$. Unlike the broad-sense heritability, which accounts for the entire genetic component of the trait, narrow-sense heritability captures only the part of the genetic component that can be approximated by a linear combination of genotyped (usually common) variants. Since the genetic component of a trait could also include non-additive genetic effects, and could also involve variants that are not captured in genotyping, it generally holds that $h^2 < H^2$.

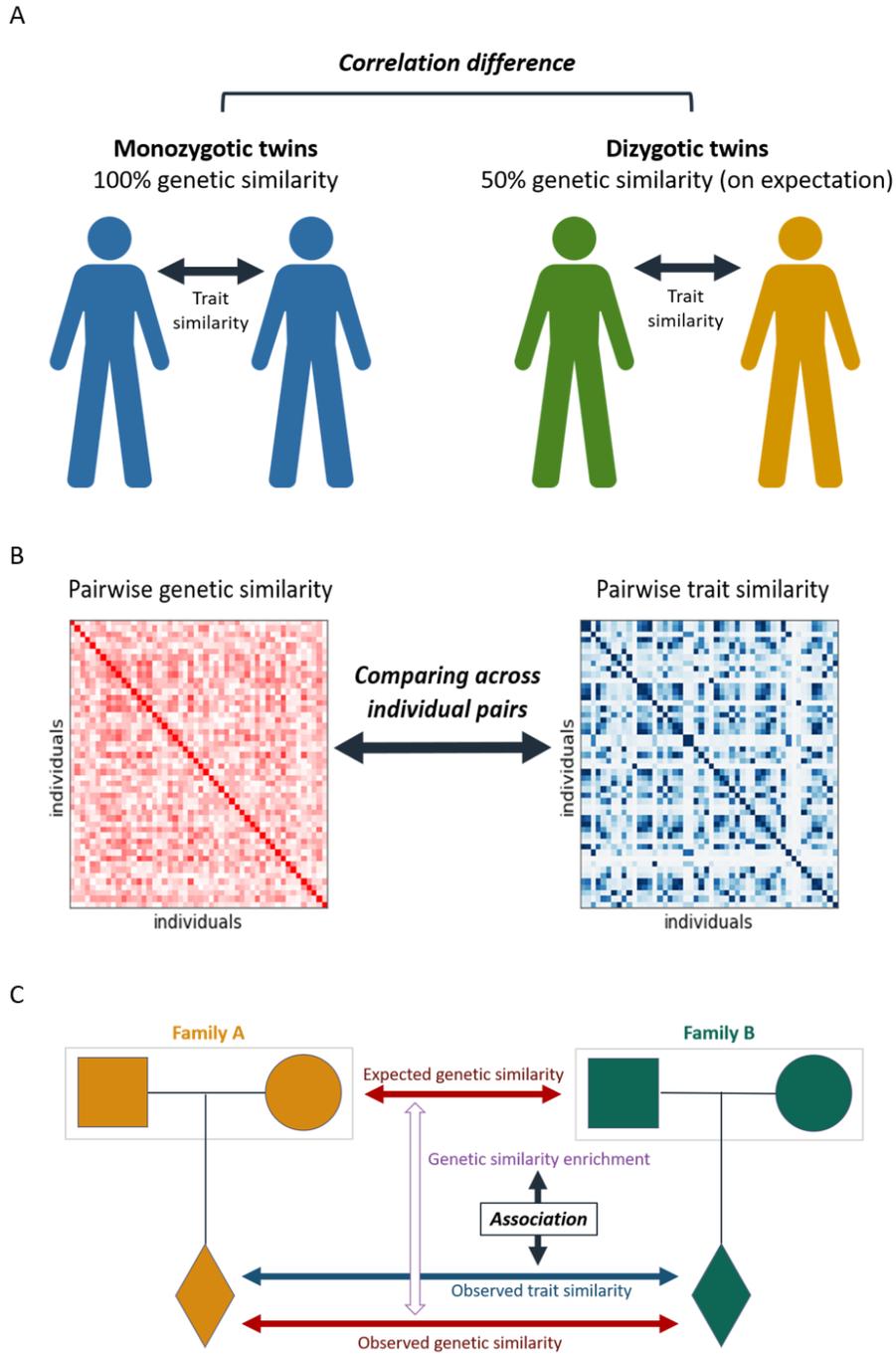

**Fig. 3: Estimating heritability**

Common methods for estimating the heritability of human traits. (**A**) In twin studies, heritability is estimated by the degree to which monozygotic (identical) twins are more phenotypically similar to each other than dizygotic (non-identical) twins. (**B**) In GREML, heritability is estimated by comparing genetic and phenotypic similarities across pairs of unrelated individuals. (**C**) In family-based methods, given a pair of individuals and their parents, the degree to which they are more genetically similar than would be expected from their parents can be compared to their phenotypic similarity to estimate the heritability of the trait.

## Interpreting heritability

In the presence of so many methods for estimating heritability, each defining heritability somewhat differently and making specific assumptions about the genetic architecture of human traits, there is an ongoing debate on which methods should be considered the most reliable and how to interpret the numbers they produce (**open problem #10: heritability estimate interpretation**). The problem is aggravated by the sensitivity of the methods to modeling assumptions and the fact that different methods often provide different estimates [62, 63], even for methods of the same family [64, 65]. For example, estimates of heritability may range from 45% to 80% for height, 23-57% for total cholesterol levels and 17-43% for educational attainment (i.e., number of years of education), depending on the method and population studied (and whether broad-sense or narrow-sense heritability is sought) [62].

As mentioned, one of the motivations to estimate heritability is getting a sense of how much predictive power we can expect from PRS. But does it mean that a trait's heritability is an absolute upper bound on the performance of a PRS? Since heritability estimation methods are never completely assumption-free, it would probably be a mistake to interpret these numbers as strict upper bounds that could never be exceeded. Furthermore, population structure and indirect genetic effects may allow genetics to capture phenotypic variance above $H^2$. Rather than a strict upper bound, it is probably more sensible to see heritability estimates as a crude assessment for the theoretical limit of PRS, given our current knowledge.

## The missing heritability problem

It was noticed early on that discovered genetic associations tended to explain only a small fraction of the heritability of most complex traits. The phenomenon was named "the missing heritability problem", expressing confusion about where most of the heritability was hiding [66]. Later, the focus of the problem has somewhat shifted, and nowadays the term "missing heritability" more commonly refers to the gap between heritability estimates from genotype and twin data [21]. Formally, it is generally the case that $h^2_{PRS} \ll h^2_{SNP} \ll h^2_{twin}$, where $h^2_{SNP}$ and $h^2_{twin}$ are the heritability estimates obtained from genotype- and twin-based methods, respectively, and $h^2_{PRS}$ is the fraction of the phenotypic variance explained by a concrete PRS. Both of these inequalities reflect possible gaps in our understanding of the genetics of human traits (**open problem #11: missing heritability**). The first gap ($h^2_{PRS} \ll h^2_{SNP}$) is more easily explained by simple statistical considerations, specifically the fact that many complex traits are highly polygenic and involve mostly weak effects (or strong but rare effects), meaning we might not have sufficient statistical power to capture the entire SNP heritability with GWAS and PRS (although this may change with greater sample sizes). The second gap ($h^2_{SNP} \ll h^2_{twin}$) necessarily reflects a failure of at least one of the two heritability estimation approaches to capture the true (broad-sense) heritability $H^2$.

Some explanations for the $h^2_{SNP} \ll h^2_{twin}$ gap argue for overestimation of the true heritability by twin studies, due to genetic interactions creating "phantom heritability" or mistaking of the effects of greater environmental similarities between monozygotic twins for genetic effects [21, 63]. Other theories argue for underestimation by SNP heritability (namely for a gap between the narrow- and broad-sense heritability). The most prominent of these theories sees rare, ungenotyped variants as the main source of missing heritability. Other explanations point out that SNP heritability could also be biased due to residual population structure [16, 21], environmental biases [62], indirect genetic effects [40] and genetic interactions (discussed later).

Some go as far as arguing that the whole notions of heritability and missing heritability are ill-posed, and that these statistical models are based on too many assumptions to be taken at face value [67]. Under this view, it could be more productive to forsake heritability estimates at this point (or at least to treat them as qualitative rather than quantitative assessments), and focus instead on improving predictions.

## Rare variants

Perhaps the most discussed potential source of missing heritability is rare variants. Variants exerting strong phenotypic effects are expected to be under intense selective pressure, and therefore remain at low frequency. As a result, genetic effects on complex traits are usually constrained to common variants exerting weak effects or rare variants with strong effects, both are statistically hard to detect and quantify.

After analyzing very large cohorts of whole-genome sequencing (which have only become available in recent years), some argue that most of the heritability can now be explained and that the missing heritability problem should be considered resolved [68], but this is still highly controversial [21] and some works argue that rare variants have overall limited contribution to heritability [69]. In inbred mice, where the allele frequency of all variants becomes either 0% or close to 50% regardless of how rare they were in the original population (thereby breaking the natural negative correlation between allele frequency and effect size), genetic associations were shown to explain a much larger fraction of the phenotypic variance, suggesting that the overall contribution of rare variants may be substantial [70]. Whether or not rare variants alone are the main source of missing heritability, by now there is a lot of evidence that they play an important role in many complex traits [71]. Unfortunately, genetic studies still rarely deal with rare variants, and there is shortage of whole-genome and whole-exome sequencing cohorts that can capture them (**open problem #4: rare variants**).

A primary problem in dealing with rare variants is that sequencing is still considerably more expensive than SNP-array genotyping, and reduction in sequencing costs has stagnated in recent years [18]. As a result, there is a real trade-off between the quality and quantity of genetic data, and it is not entirely clear which of the two is more critical. The recent release of 200K whole-exome sequences by the UK Biobank [72], in addition to SNP-array genotypes over the same individuals, could provide valuable insight into the cost-effectiveness of different types of genetic data. Another problem is that even after a large cohort is fully sequenced, it is not entirely clear how to interpret rare variants. Association studies and PRS typically require variants to be frequent enough to accurately assess their effects. Dealing with rare variants may require more sophisticated methods that look beyond the statistical patterns of specific variants. A promising class of methods is burden tests, which consider the aggregated effects of multiple variants sharing the same gene or genomic locus [33–37, 71]. Another approach to analyzing the effects of rare variants would be to leverage family studies [20]. Additional problems with rare variants are that they are more sensitive to residual population structure [21] and quality control (as it is harder to distinguish true variants from sequencing errors in the presence of limited data).

# Non-additive genetic effects: oversight or non-issue?

## The additive genetic model

Nearly all methods for studying the genetics of complex human traits, from GWAS to PRS to heritability estimates, assume additivity in the effects of variants. The typical additive model for a continuous trait is $y = \beta_1 g_1 + \ldots + \beta_k g_k + \epsilon$, where $y$ is an individual's phenotype value, $g_1, \ldots, g_k$ are the individual's genotype values over $k$ variants, $\beta_1, \ldots, \beta_k$ are constant per-variant weights indicating the effect size of each of the $k$ variants, and $\epsilon$ is Gaussian noise (typically assumed to be independent across individuals) capturing the remainder of the phenotype (including environmental, non-additive and random effects, as well as the genetic effects of unincluded variants). When dealing with a binary or categorical trait (e.g., a disease diagnosis), the model is typically altered into $P(y = 1) = \sigma(\beta_1 g_1 + \ldots + \beta_k g_k)$, where $\sigma(x) = exp(x)/(exp(x) + 1)$ is the standard logistic function, and $P(y = 1)$ is the probability of the individual to have the trait. It is also common to include in the model other covariates in addition to the genotype values, such as sex, age and the principal components of genetic variation (to account for population structure). Consistent with the overall assumption of additivity, the genotype values $g_1, \ldots, g_k$ are typically also encoded in an additive way, namely each $g_i$ is set as 0, 1 or 2, depending on the number of copies of the alternative allele of variant $i$ that the individual has. Following the 0/1/2 encoding, genotype values are commonly normalized such that each $g_i$ would have a mean of 0 and standard deviation of 1 over the study cohort.

Within this framework, $y$ and $g_1, \ldots, g_k$ are observed, and the study's goal is to make inference about the coefficients $\beta_1, \ldots, \beta_k$. In GWAS, one is typically interested in finding out whether $\beta_i \neq 0$ for each of the tested variants by calculating a p-value for the null hypothesis $H_0: \beta_i = 0$. In PRS, the goal is to simultaneously estimate all of the coefficients $\beta_1, \ldots, \beta_k$ and to use the expression $\beta_1 g_1 + \ldots + \beta_k g_k$ as a predictor for the phenotype $y$. In heritability estimation, the objective is to estimate the overall genetic variance $Var(\beta_1 g_1 + \ldots + \beta_k g_k)$.

## Is the additive model (more or less) true?

It is obvious to anyone familiar with complex biological systems that genetic effects are not truly linear. Nonetheless, many argue that linear genetic models are a good-enough approximation of the real biological complexity [73]. According to this common view, the contribution of non-additive genetic effects to the variance of most traits is low (i.e., additive genetic effects account for most of the heritability), so we need not worry too much about them. By now the additive genetic model has become so mainstream that it is commonly just taken for granted without any explicit justification. Despite its popularity and despite being very convenient for computational and statistical analysis, it is important to understand the empirical and theoretical support in favor of the additive genetic model, and consider the possibility that it might nonetheless be wrong or incomplete.

A strong empirical result in favor of the additive genetic model was presented in a 2015 meta-analysis covering virtually all twin studies published between 1958 and 2012 [59]. It was found that, across many different traits, the phenotypic correlation between monozygotic twins was roughly twice the correlation between dizygotic twins, consistent with a model of mostly additive genetic effect and no shared environmental influences on twins. Specifically, 69% of the 2,748 analyzed twin studies were consistent with the null hypothesis $r_{MZ} = 2r_{DZ}$. However, results in this meta-analysis were shown only for groups of traits (such as "cognitive traits"), but not for individual phenotypes. Furthermore, the heritability of some traits (such as height) had been studied much more extensively than others, and most of the analyzed studies had investigated continuous (rather than binary or categorical) traits. Another limitation acknowledged by the authors was the presence of substantial overlap in the twins recruited across the different studies.

Also supporting the additive model is the poor track record of nonlinear phenotype prediction models, which generally do not substantially outperform linear PRS. On the other hand, there haven't been that many attempts to develop such models, and it might be too early to give up on outperforming linear PRS (see next section).

Another argument in favor of additivity appeals to theoretical considerations that, under certain assumptions and in the presence of a sufficiently large number of causal variants (i.e., if the trait is sufficiently polygenic), the overall phenotypic variance attributed to additive genetic effects will almost certainly dominate the epistatic (i.e., non-additive) genetic effects [74]. On the other hand, a recent simulation analysis of phenotypes with deep neural networks showed that nonlinear phenotypes (that cannot be approximated well by a linear model) are possible [75].

An empirical evidence against the additive model is the fact that epistatic genetic effects are often detected when sought, and possibly contribute to phenotypic variance quite substantially in some non-human organisms (see later section). Another argument opposing the additive genetic model is that we still have fundamental gaps in our understanding of the genetic architectures of complex human traits, and the assumption of additivity (made by virtually all heritability estimation methods) should be considered an immediate suspect for the missing heritability problem [63, 67]. Given what we know about complex biological systems, the burden of proof should be on those arguing for additive genetic effects being the primary source of heritability.

It is also important to note that a lot of the research and discussion on the genetic architecture of complex human traits, arguably too much of it, has focused on a rather small set of human traits, and especially on human height. It is possible that we have been somewhat led astray by focusing too much on a non-representative example. Height is actually rather unusual within the human phenotypic landscape in how heritable it is, and it is also very polygenic (although there are more polygenic traits [26]). Perhaps it is also uncommonly additive. It is also possible that continuous traits, whose genetic architecture is easier to study, behave differently than diseases and other binary traits.

Despite many years of debate on the question, we argue that the question of additivity is still not fully settled and that more work is needed to determine whether non-additive genetic effects underly substantial phenotypic variance in complex human traits (**open problem #2: non-additive and epistatic genetic effects**). It would also benefit the discussion on additivity if the claims being defended or argued against were more precisely defined. For example, there is a huge difference between arguing that over 50% of the phenotypic variance attributed to genetics could be approximated by a linear model, which is a rather modest claim, and arguing for near 100%, which is a much stronger claim. We would also like to see evidence in favor or against genetic additivity being presented over more diverse human phenotypes and with more clarity and rigor, in particular with respect to the modeling assumptions underlying such works. When reporting or interpreting the results of genetic studies, it is important to be mindful of the assumption of additivity.

## Consequences of non-additivity to heritability and PRS

The question of additivity is tightly related to the question of missing heritability (open problem #11: missing heritability), and is therefore also relevant to the performance of PRS (open problem #14: genotype-to-phenotype prediction performance). As virtually all contemporary heritability estimation methods assume additivity, non-additive genetic effects could be part of the explanation for the missing heritability problem. As mentioned, SNP heritability does not include non-additive variance, so it is expected that methods such as GREML and LDSC would underestimate the full (broad-sense) heritability. It was demonstrated through simulation analysis that methods for SNP heritability estimation may dramatically underestimate the heritability of nonlinear phenotypes (and that the more nonlinear the simulated phenotypes are, the less heritability is recovered) [75]. Twin studies, on the other hand, assume that the correlation between the genetic effects of non-identical twins is exactly half, which holds for additive but not epistatic effects. As a result, twin studies are likely to overestimate the true heritability in the presence of epistasis [63]. Part of the gap between $h^2_{SNP}$ and $h^2_{twin}$ could therefore be attributed, at least in principle, to non-additivity.

If non-additive effects indeed make up a big part of the heritability of complex traits, then this has important ramifications for PRS: we would expect linear models to be limited, at least in principle, compared to the full potential of nonlinear models. Some attempts have been made to train PRS with nonlinear models, including support vector machines, random forests and deep neural networks [76–78]. These nonlinear algorithms have a strong track record across numerous domains of machine learning. However, in the context of PRS, such attempts have generally failed to outperform simple linear models, providing strong evidence in favor of the theory stating that nonlinear effects are not that important in the grand scheme of things, but these indications are not yet decisive. Attempts to model genetic effects in a nonlinear way have only been made a handful of times, and there could be other reasons for off-the-shelf machine-learning algorithms having a hard time to pick nonlinear genetic effects. For example, we know that in genetics, to a much greater extent than in other domains where machine learning is commonly applied, data is often very noisy, effects are very weak (or rare), and there doesn't exist an obvious and easily-exploitable structure to the data. It would be interesting to see if future works are able to capture a substantial non-additive genetic signal.

The adequate performance of linear models compared to nonlinear alternatives is the primary reason for their dominance in genetic studies, but there are also other motivations swaying researchers to favor them. Other advantages of linear models include high robustness and interpretability, both are crucial for clinical applicability. Also of great importance to researchers is the capacity to work with summary statistics of published results without being dependent on individual-level data. This is easily done with linear models, but usually impossible with nonlinear algorithms. Since human genetic datasets with individual-level information tend to be highly restricted (due to privacy concerns), this often becomes a critical consideration. If large-scale, highly accessible biobanks (such as the UK Biobank) become more prevailing in the coming years, we might see nonlinear methods becoming more popular.

# Epistatic, dominant and recessive effects in complex traits

Related to the principal question of non-additivity is the more practical problem of finding epistatic genetic interactions (known as GxG), namely finding combinations (usually pairs) of variants or genes that interact together in a nonlinear way [79]. This is a notoriously difficult combinatorial and computational problem [4]. For example, scanning a dataset with a million genotyped variants for all possible pairwise interactions would involve half a trillion (~5E11) variant pairs. Obtaining sufficient statistical power to find significant pairs under such conditions would require much stronger effects than those required to find additive genetic effects. Even if additive models are in fact capable of capturing most of the genetic component of phenotypic variance, finding epistatic effects might still be important for understanding the mechanism of human traits.

A lot of epistasis research has focused on non-human model organisms, where the phenomenon could be studied with experiments (including genome screening projects with libraries of double gene deletions) [4]. Specifically, a lot of research has revolved around yeast. The general conclusion from these studies is that epistasis is indeed quite prevalent in non-human organisms and could contribute quite substantially to the phenotypic variance [80, 81]. However, many of the experimental studies involve artificially-induced mutations or inbred populations and therefore do not provide direct evidence on the scale of the phenomenon with respect to natural genetic variation.

Recessive inheritance is an interesting special case of non-additive genetic interactions. While dominant and recessive inheritance play an important role in the study of Mendelian traits, these inheritance modes are hardly studied in the context of complex traits. In principle, dominant and recessive genetic effects are perfectly applicable to complex traits as well as Mendelian. It should be noted that the textbook definitions of dominant and recessive effects are often just approximations of the genetic effects found in the real world. For example, there are Mendelian diseases exhibiting imperfect recessive inheritance, as in the case of thalassemia (a type of anemia caused by mutations in the hemoglobin genes), which is considered a recessive disorder, but individuals who carry one copy of a deleterious mutation may develop mild symptoms of anemia. Since strong genetic effects are mostly confined to rare variants, an additive model is usually a good approximation for dominant effects but a very poor approximation for recessive effects. The additive model mentioned earlier can in principle accommodate dominant and recessive effects at the variant level through changing of the encoding of the genotype values $g_i$. Specifically, assigning the values 0/0/1 or 0/1/1 instead of 0/1/2 would capture recessive or dominant effects at the variant level, respectively. However, it is anticipated that many recessive effects would occur at the gene level and not at the variant level. If the two copies of a gene are affected by different variants, a case known as compound heterozygosity, then a variant-level recessive model would be completely blind to it. Due to the inherent blindness of GWAS to compound heterozygosity, the study of recessive genetic effects in complex human traits is highly neglected, but some recent works show that they are in fact common in complex traits [82, 83]. Gene-based methods (as opposed to variant-based methods) seem especially promising for detecting recessive effects [37].

# Acknowledging the complexity of genetics

## Gene-environment interactions

Similar to epistatic effects (GxG), interactions between genetic and environmental factors, known as GxE, also play a role in complex human traits [84, 85]. GxE interactions indicate that the effect of a genetic factor is dependent on the presence of an environmental factor (or, equivalently, that the effect of the environmental factor is dependent on the genetic factor), where "environmental factor" means any non-genetic determinant of the trait (including epigenetics or even random conditions). An obvious example is the effect of sunlight exposure (an environmental factor) on melanoma (a phenotype) being dependent on the genetic variants that determine skin color. A less obvious example is different levels of colorectal cancer and adenoma risks associated with the missense variant A222V in the MTHFR gene (methylenetetrahydrofolate reductase) depending on the levels of folic acid dietary intake [84]. Various statistical methods and computational approaches have been developed to detect GxE interactions [86].

There are many pragmatic reasons to be interested in GxE. By taking into account the modifying effects of environmental factors, more genetic associations could be found [85]. Furthermore, gene-environment interactions can provide insight into the mechanism of genetic associations (open problem #13: from causality to mechanism). Additionally, when GxG or GxE interactions exist, it is possible that genetic associations and PRS will not perform uniformly across human groups (in particular across space and time). Specifically, it might be possible to identify subgroups showing unusually high risk for genetic or environmental factors. As a result, GxE interactions could allow individualized genetic-based interventions by modifying the interacting environmental factors (thereby reducing disease risk) [85]. Special interest in GxE is present in the social sciences (especially in psychology), where gene-environment interactions are believed to play a central role in human personality and behavior, and an active debate is still going around the old question of "nature versus nurture".

While generally recognized as an important piece in the genetic puzzle of complex human traits, GxE is a notoriously difficult subject of study, and there has been limited progress in addressing it (**open problem #3: gene-environment interactions**) [86]. It is very challenging to run reliable and robust GxE studies. For example, it is difficult to properly address potential confounding, even more than in standard genetic studies [86]. Over the years there have been few concrete GxE discoveries, and even fewer successful replications. Discovering GxE interactions is hard in part due to the vast combinatorial search space (as in GxG) [84]. When the genetic variants and environmental conditions under investigation are both rare, detecting an interaction between them becomes nearly impossible. On top of all the regular challenges in replicating genetic associations, GxE replication studies and meta-analyses are often hindered by differences in how environmental factors are defined and measured across studies [85]. Even rigorously defining what exactly GxE means is not trivial, as it requires to describe what the architecture of a trait should look like in the absence of interactions (when both genetics and the environment may influence the trait, but separately). Specifically, the presence or absence of GxE could depend on the scaling of the phenotype. In the case of a binary phenotypic outcome and binary genetic and environmental factors, it would depend on whether we expect risk for the outcome to depend additively or multiplicatively on the separate risks associated with the genetic and environmental factors [84]. When the interacting environmental factors are themselves affected by genetic factors, GxE interactions could be mediators of underlying GxG interactions, meaning that GxG and GxE are often tightly related.

The study of GxE has important ramifications for the concept of heritability and the question of missing heritability. There have been works trying to quantify the overall contribution of GxE to phenotypic variance (for certain collections of environmental factors, and under certain assumptions about the genetic architecture of the studied traits). These attempts have often ended up with the conclusion that GxE interactions contribute quite substantially to the phenotypic variance of some traits (such as BMI and pulse pressure) [87, 88], yet the overall importance of GxE to trait heritability is still controversial [73, 86]. Notably, the mere definition of heritability is tricky in the presence of GxE interactions. To define heritability, continuous phenotypes are commonly decomposed as $P = G + E$, where $P$ is the phenotype value, $G$ is the overall contribution of genetic factors to the phenotype, and $E$ is the contribution of environmental factors (defined as the residual term encapsulating all non-genetic factors contributing to the trait). It then holds that $Var(P) = Var(G) + Var(E) + 2Cov(G,E)$. This means that when $Cov(G,E) \neq 0$ (in the presence of GxE interactions), $Var(G)$ will not capture all the variance of the trait that is due to genetics, as implicitly assumed by defining heritability as $H^2 = Var(G)/Var(P)$ (and if $Cov(G,E)$ is negative, it may even capture too much). It has been argued that environmental effects, when not properly accounted for, may indeed inflate heritability estimates. For example, it has been asserted that heritability estimates for educational attainment might be inflated by ~70% [62]. Some have gone as far as arguing that just as we cannot say how much of the area of a rectangle is due, separately, to each of its two dimensions, so it is not possible to separate "nature" from "nurture" [67].

## Selection bias

Selection bias describes a situation where a study cohort does not perfectly represent the population that it is presumed to represent, leading to unjustified conclusions about the general population. Since participation in human studies is generally voluntary, this is a difficult challenge to overcome. While the problem and the challenges it presents are by no means unique to genetic studies, human genetics researchers appear to have been surprisingly unconcerned about the implications of selection bias. Only very recently did the problem gain some attention, following a few demonstrations of its potential to lead to false discoveries. For example, a recent study detected statistically significant associations between autosomal variants and sex [89]. Since such associations are clearly spurious (autosomal chromosomes cannot affect sex), it was concluded to be the result of participation bias. Specifically, if both sex and a particular genetic variant affect an individual's decision to participate in the study cohort (which can be seen as a behavioral trait), then we would have a collider bias when conditioning on that decision. Another special case of selection bias is survival bias. If we recruit individuals with a history of cancer to participate in a genetic study, then it could be biased by individuals with extremely aggressive tumors not having survived to participate. It was also demonstrated that the common case-control study design (where cases and controls are recruited separately) can lead to underestimation of SNP heritability, thereby adding to the missing heritability [90]. Since the problem of selection bias in genetic studies was acknowledged only recently, not enough work has been dedicated to fully understanding the scope of the problem and working out methods and best practices to minimize its effect (**open problem #9: selection bias**).

## Overlooked genetic variation

We have deliberately focused this review on problems that are specific to the link between genotype and phenotype, putting aside most of the upstream technical aspects of collecting genetic data (such as issues related to sequencing, variant calling and quality control). Nonetheless, it is crucial to acknowledge that, like in any field, the results of our research will only be as good as the quality of the data we use. In particular, there are certain types of genetic variation that are systematically underrepresented, or altogether absent, from contemporary genetic studies (**open problem #5: non-standard genetic variation**).

We have already touched on the challenges of studying the effects of rare variants due to data scarcity (open problem #4: rare variants). Other neglected variants are those in sex chromosomes (X and Y) [82], which are often excluded from genetic studies due to mapping and variant calling challenges, and also because they diverge from diploidy and, as a result, from common modeling assumptions. The mitochondrial chromosome (MT) presents similar challenges, and is even more commonly left out. Structural and copy-number variations also pose a major challenge to existing protocols and tend to be ignored, as they are hard to genotype and sometimes also deviate from standard modeling [66, 91, 92]. For example, it is open to interpretation what dominant or recessive inheritance would mean in the case of copy number variation. Genetic variation in repetitive regions of the genome also tends to be ignored due to mapping challenges [93]. Part of the gap between heritability estimates from twin and genotype data could be due to under-genotyped genetic variation [21, 66, 67]. Advanced genotyping technologies such as nanopore and PacBio sequencing may play a role in resolving complex genetic variants [94, 95].

Another aspect of genetic variation that tends to be overlooked is phasing, namely whether two heterozygous variants occur on the same or different copies of the chromosome (and their maternal or paternal origin) [96]. This could be important, for example, when dealing with compound heterozygosity (where a recessive effect takes place only when variants occur at both copies of a gene) or epistatic effects involving cis-regulatory elements. Another aspect of human genetic variation regularly neglected is mosaicism, namely the occasional occurrence of mutational events during cell division leading to genetic diversity between cells in the body (with more cells having a genetic alteration the earlier it occurred during development) [97]. Genetic studies assume that an individual's genetic sample represents their entire genetic repertoire, but that is not perfectly accurate. There is evidence suggesting that genetic mosaics could have phenotypic effects in the brain in traits such as autism spectrum disorder [97]. Whether this is an anecdotal phenomenon or an important aspect of human genetics still remains to be determined. Should mosaicism be taken into account in genetic studies, or is it better treated as yet another non-genetic factor (like other omics and environmental effects)?

A related question is how to represent genetic variation. The accepted norm today is to represent an individual's genome as a set of unrelated variants indicating deviation from the human reference genome (which is itself a somewhat artificial and incomplete construct [98]). Is it necessarily the best way to represent genetic variation and study its effects on human traits? A recently explored alternative is to skip variant calling altogether and to seek direct associations of the trait with raw sequencing reads [99]. Another long-standing idea is to represent genomes as graphs of haplotypes [100].

In a sense we have only been dealing with the easiest parts of human genetics, focusing on common, simple autosomal variants, without genetic and environmental interactions. The overall contribution of unaccounted genetic variation to heritability, and whether this is a major source of missing heritability, is yet to be determined.

## Phenotype definition

Just as the quality and representation of genotypes is expected to have a major influence on the results of genetic studies, so do the technicalities of how human phenotypes are defined and measured. While some phenotypes are straightforward to measure accurately (e.g., height), other traits are a lot more nuanced and could be defined in different ways. For example, there has been a long discussion on whether schizophrenia and bipolar disorder, two distinct but highly related psychiatric diagnoses with strongly overlapping genetic determinants, should be seen as residing on a single psychosis continuum [101]. Moreover, different physicians may end up assigning the same individual with different diagnoses [102], and some traits are prone to diagnostic errors [103]. Diagnostic protocols may also vary between health or insurance systems, even within the same country. A related question is whether to define disease status based on official clinical diagnoses or self-reports [46]. Ill-defined phenotypes are expected to introduce random noise and systematic biases into genetic studies [104].

It has been postulated that some clinical diagnoses are in fact a combination of distinct biological phenotypes, each involving separate genetic and cellular pathways, and that the similar clinical manifestations of these different phenotypes are only superficial. In such cases, it could be beneficial to study each of the subphenotypes separately [105]. Conversely, it is also possible that some diagnoses that are considered distinct in fact share a common biological etiology and are therefore better defined as a single phenotype. There are some preliminary efforts to use genetics as a guideline for better phenotype definitions [106]. Improving the methodology and practices used to define and measure phenotypes would be highly beneficial (**open problem #8: phenotype definition**).

Sex differences, which are prevalent across the human phenome, could be seen as an important special case of subphenotypes [105]. When sex interacts with genetics in a non-additive way, sex differences can also be seen as a special case of GxE and GxG interactions (in this case, the interaction of sex chromosomes with other genetic factors). The common practice today in GWAS and other genetic studies is to treat sex as yet another covariate that has to be accounted for. But given its profound effect on so many human traits, perhaps more comprehensive approaches are needed. For example, it might be a constructive norm to study and report on males and females separately, when sample sizes and statistical power allow that.

# Open discussion on open problems

We have attempted to provide a broad view on current important open problems in the field of human trait genetics. Towards this goal, we reached out and spoke with many active researchers in the field covering a diverse set of backgrounds and perspectives. Following these conversations, we converged on a list of 16 open problems that were consistently acknowledged by us and others to be particularly important (Table 1 and Supplementary Table S1). Our main criterion for deeming an open problem sufficiently important was: would it considerably help realizing the field's potential if this problem were solved?

We attempted to provide an up-to-date view on the field, focusing on the open problems that appear to constrain the field in the near future. Despite the efforts we have made, we are aware that this review and our list of open problems are far from being truly exhaustive. There are many additional relevant open problems that we judged to be relatively minor, and it is also very plausible that we have overlooked some major open problems.

We hope that this presentation of open problems will contribute to the open discussion already taking place. In particular, critical discussion on our list of open problems (and on open problems we have overlooked) would be beneficial. We think that an open dialogue is especially important among researchers and practitioners from diverse backgrounds and expertise, for example between researchers and clinicians. Other important discussions should include technology providers (e.g., of high-throughput sequencing and data processing) and resource providers (such as biobanks). For example, whether data resources are dominated by publicly-available biobanks or restricted databases will greatly affect the kind of challenges we face.

# Acknowledgements

We would like to thank Alexander (Sasha) Gusev, Alexander l. Young, Noah Zaitlen, Or Zuk, Sagiv Shifman, Shai Ben-Shahar, Shai Carmi, Shiri Shakedi, Regev Schweiger and Yossi Farjoun for offering us their perspectives on open problems in the field. This review does not necessarily reflect their views, and all mistakes are ours.

We also thank Vladimir Gritsenko and Yedael Y. Waldman for valuable feedback on our initial draft.

# Supplementary materials

**Supplementary Table S1**: The list of 16 open problems, with details about each problem. [ftp.cs.huji.ac.il/users/michall/open_problems_in_human_trait_genetics_preprint/Supplementary_Table_S1.xlsx]

# References

1. Lander ES (1996) The new genomics: global views of biology. Science (80- ) 274:536–539
2. Visscher PM, Brown MA, McCarthy MI, Yang J (2012) Five years of GWAS discovery. Am J Hum Genet 90:7–24
3. Visscher PM, Wray NR, Zhang Q, et al (2017) 10 years of GWAS discovery: biology, function, and translation. Am J Hum Genet 101:5–22
4. Tam V, Patel N, Turcotte M, et al (2019) Benefits and limitations of genome-wide association studies. Nat Rev Genet 20:467–484
5. Choi SW, Mak TS-H, O'Reilly PF (2020) Tutorial: a guide to performing polygenic risk score analyses. Nat Protoc 15:2759–2772. https://doi.org/10.1038/s41596-020-0353-1
6. Visscher PM, Hill WG, Wray NR (2008) Heritability in the genomics era—concepts and misconceptions. Nat Rev Genet 9:255–266
7. Torkamani A, Wineinger NE, Topol EJ (2018) The personal and clinical utility of polygenic risk scores. Nat Rev Genet 19:581
8. Davey Smith G, Hemani G (2014) Mendelian randomization: genetic anchors for causal inference in epidemiological studies. Hum Mol Genet 23:R89–R98. https://doi.org/10.1093/hmg/ddu328
9. Hamosh A, Scott AF, Amberger JS, et al (2005) Online Mendelian Inheritance in Man (OMIM), a knowledgebase of human genes and genetic disorders. Nucleic Acids Res 33:D514--D517
10. Robinson MR, Kleinman A, Graff M, et al (2017) Genetic evidence of assortative mating in humans. Nat Hum Behav 1:1–13
11. Cardon LR, Palmer LJ (2003) Population stratification and spurious allelic association. Lancet 361:598–604
12. Price AL, Patterson NJ, Plenge RM, et al (2006) Principal components analysis corrects for stratification in genome-wide association studies. Nat Genet 38:904–909
13. Peterson RE, Kuchenbaecker K, Walters RK, et al (2019) Genome-wide association studies in ancestrally diverse populations: opportunities, methods, pitfalls, and recommendations. Cell 179:589–603
14. Yang J, Zaitlen NA, Goddard ME, et al (2014) Advantages and pitfalls in the application of mixed-model association methods. Nat Genet 46:100–106. https://doi.org/10.1038/ng.2876
15. Mbatchou J, Barnard L, Backman J, et al (2021)

Computationally efficient whole-genome regression for quantitative and binary traits. Nat Genet 53:1097–1103. https://doi.org/10.1038/s41588-021-00870-7

16. Abdellaoui A, Verweij KJH, Nivard MG (2021) Geographic confounding in genome-wide association studies. bioRxiv

17. Sohail M, Maier RM, Ganna A, et al (2019) Polygenic adaptation on height is overestimated due to uncorrected stratification in genome-wide association studies. Elife 8:e39702

18. McGuire AL, Gabriel S, Tishkoff SA, et al (2020) The road ahead in genetics and genomics. Nat Rev Genet 1–16

19. Laird NM, Lange C (2006) Family-based designs in the age of large-scale gene-association studies. Nat Rev Genet 7:385–394

20. Wijsman EM (2012) The role of large pedigrees in an era of high-throughput sequencing. Hum Genet 131:1555–1563

21. Young AI (2019) Solving the missing heritability problem. PLoS Genet 15:e1008222

22. Sudlow C, Gallacher J, Allen N, et al (2015) UK biobank: an open access resource for identifying the causes of a wide range of complex diseases of middle and old age. PLoS Med 12:e1001779

23. Bycroft C, Freeman C, Petkova D, et al (2017) Genome-wide genetic data on~ 500,000 UK Biobank participants. BioRxiv 166298

24. Feliciano P, Daniels AM, Snyder LG, et al (2018) SPARK: a US cohort of 50,000 families to accelerate autism research. Neuron 97:488–493

25. Consortium HR, others (2016) A reference panel of 64,976 haplotypes for genotype imputation. Nat Genet 48:1279–1283

26. Weissbrod O, Hormozdiari F, Benner C, et al (2020) Functionally informed fine-mapping and polygenic localization of complex trait heritability. Nat Genet 1–9

27. Pasaniuc B, Price AL (2017) Dissecting the genetics of complex traits using summary association statistics. Nat Rev Genet 18:117–127. https://doi.org/10.1038/nrg.2016.142

28. Fadista J, Manning AK, Florez JC, Groop L (2016) The (in) famous GWAS P-value threshold revisited and updated for low-frequency variants. Eur J Hum Genet 24:1202–1205

29. Spain SL, Barrett JC (2015) Strategies for fine-mapping complex traits. Hum Mol Genet 24:R111--R119

30. Kichaev G, Yang W-Y, Lindstrom S, et al (2014) Integrating functional data to prioritize causal variants in statistical fine-mapping studies. PLoS Genet 10:e1004722

31. Weeks EM, Ulirsch JC, Cheng NY, et al (2020) Leveraging polygenic enrichments of gene features to predict genes underlying complex traits and diseases. medRxiv

32. Richards S, Aziz N, Bale S, et al (2015) Standards and guidelines for the interpretation of sequence variants: a joint consensus recommendation of the American College of Medical Genetics and Genomics and the Association for Molecular Pathology. Genet Med 17:405–423

33. Wu MC, Lee S, Cai T, et al (2011) Rare-variant association testing for sequencing data with the sequence kernel association test. Am J Hum Genet 89:82–93

34. Lee S, Emond MJ, Bamshad MJ, et al (2012) Optimal unified approach for rare-variant association testing with application to small-sample case-control whole-exome sequencing studies. Am J Hum Genet 91:224–237

35. Gamazon ER, Wheeler HE, Shah KP, et al (2015) A gene-based association method for mapping traits using reference transcriptome data. Nat Genet 47:1091

36. Gusev A, Ko A, Shi H, et al (2016) Integrative approaches for large-scale transcriptome-wide association studies. Nat Genet 48:245

37. Brandes N, Linial N, Linial M (2020) PWAS: proteome-wide association study—linking genes and phenotypes by functional variation in proteins. Genome Biol 21:1–22

38. Kryshtafovych A, Schwede T, Topf M, et al (2019) Critical assessment of methods of protein structure prediction (CASP)—Round XIII. Proteins Struct Funct Bioinforma 87:1011–1020

39. Zhou N, Jiang Y, Bergquist TR, et al (2019) The CAFA challenge reports improved protein function prediction and new functional annotations for hundreds of genes through experimental screens. Genome Biol 20:1–23

40. Young AI, Benonisdottir S, Przeworski M, Kong A (2019) Deconstructing the sources of genotype-phenotype associations in humans. Science (80- ) 365:1396–1400

41. Amberger JS, Bocchini CA, Scott AF, Hamosh A (2019) OMIM. org: leveraging knowledge across phenotype--gene relationships. Nucleic Acids Res 47:D1038--D1043

42. Buniello A, MacArthur JAL, Cerezo M, et al (2018) The NHGRI-EBI GWAS Catalog of published

genome-wide association studies, targeted arrays and summary statistics 2019. Nucleic Acids Res 47:D1005--D1012

43. Boyle EA, Li YI, Pritchard JK (2017) An expanded view of complex traits: from polygenic to omnigenic. Cell 169:1177–1186

44. Shohat S, Amelan A, Shifman S (2020) Convergence and Divergence in the Genetics of Psychiatric Disorders from Pathways to Developmental Stages. Biol Psychiatry

45. Dixit A, Parnas O, Li B, et al (2016) Perturb-Seq: dissecting molecular circuits with scalable single-cell RNA profiling of pooled genetic screens. Cell 167:1853–1866

46. Ge T, Chen C-Y, Neale BM, et al (2017) Phenome-wide heritability analysis of the UK Biobank. PLoS Genet 13:e1006711

47. Lello L, Avery SG, Tellier L, et al (2018) Accurate genomic prediction of human height. Genetics 210:477–497

48. Moldovan A, Waldman YY, Brandes N, Linial M (2021) Body Mass Index and Birth Weight Improve Polygenic Risk Score for Type 2 Diabetes. J Pers Med 11:. https://doi.org/10.3390/jpm11060582

49. Lewis CM, Vassos E (2020) Polygenic risk scores: from research tools to clinical instruments. Genome Med 12:1–11

50. Chung W, Chen J, Turman C, et al (2019) Efficient cross-trait penalized regression increases prediction accuracy in large cohorts using secondary phenotypes. Nat Commun 10:1–11

51. Bareinboim E, Pearl J (2016) Causal inference and the data-fusion problem. Proc Natl Acad Sci 113:7345–7352

52. Daly AK (2017) Pharmacogenetics: a general review on progress to date. Br Med Bull 124:65–79

53. Gabai-Kapara E, Lahad A, Kaufman B, et al (2014) Population-based screening for breast and ovarian cancer risk due to BRCA1 and BRCA2. Proc Natl Acad Sci 111:14205–14210

54. Lencz T, Backenroth D, Granot-Hershkovitz E, et al (2021) Utility of polygenic embryo screening for disease depends on the selection strategy. bioRxiv 2011–2020

55. Turley P, Meyer MN, Wang N, et al (2021) Problems with Using Polygenic Scores to Select Embryos. N Engl J Med 385:78–86. https://doi.org/10.1056/NEJMsr2105065

56. Hilker R, Helenius D, Fagerlund B, et al (2018) Heritability of schizophrenia and schizophrenia spectrum based on the nationwide Danish twin register. Biol Psychiatry 83:492–498

57. Visscher PM, Wray NR (2015) Concepts and misconceptions about the polygenic additive model applied to disease. Hum Hered 80:165–170

58. Selita F, Kovas Y (2019) Genes and Gini: what inequality means for heritability. J Biosoc Sci 51:18–47

59. Polderman TJC, Benyamin B, De Leeuw CA, et al (2015) Meta-analysis of the heritability of human traits based on fifty years of twin studies. Nat Genet 47:702–709

60. J Mayhew A, Meyre D (2017) Assessing the heritability of complex traits in humans: methodological challenges and opportunities. Curr Genomics 18:332–340

61. Yang J, Zeng J, Goddard ME, et al (2017) Concepts, estimation and interpretation of SNP-based heritability. Nat Genet 49:1304

62. Young AI, Frigge ML, Gudbjartsson DF, et al (2018) Relatedness disequilibrium regression estimates heritability without environmental bias. Nat Genet 50:1304–1310. https://doi.org/10.1038/s41588-018-0178-9

63. Zuk O, Hechter E, Sunyaev SR, Lander ES (2012) The mystery of missing heritability: Genetic interactions create phantom heritability. Proc Natl Acad Sci 109:1193–1198

64. Speed D, Holmes J, Balding DJ (2020) Evaluating and improving heritability models using summary statistics. Nat Genet 52:458–462. https://doi.org/10.1038/s41588-020-0600-y

65. Speed D, Cai N, Johnson MR, et al (2017) Reevaluation of SNP heritability in complex human traits. Nat Genet 49:986

66. Manolio TA, Collins FS, Cox NJ, et al (2009) Finding the missing heritability of complex diseases. Nature 461:747

67. Génin E (2020) Missing heritability of complex diseases: case solved? Hum Genet 139:103–113. https://doi.org/10.1007/s00439-019-02034-4

68. Wainschtein P, Jain DP, Yengo L, et al (2019) Recovery of trait heritability from whole genome sequence data. BioRxiv 588020

69. Schoech AP, Jordan DM, Loh P-R, et al (2019) Quantification of frequency-dependent genetic architectures in 25 UK Biobank traits reveals action of negative selection. Nat Commun 10:1–10

70. Flint J, Eskin E (2012) Genome-wide association studies in mice. Nat Rev Genet 13:807–817

71. Wang Q, Dhindsa RS, Carss K, et al (2021) Rare variant contribution to human disease in 281,104 UK

Biobank exomes. Nature. https://doi.org/10.1038/s41586-021-03855-y

72. Szustakowski JD, Balasubramanian S, Kvikstad E, et al (2021) Advancing human genetics research and drug discovery through exome sequencing of the UK Biobank. Nat Genet 53:942–948. https://doi.org/10.1038/s41588-021-00885-0
73. Hill WG, Goddard ME, Visscher PM (2008) Data and theory point to mainly additive genetic variance for complex traits. PLoS Genet 4:e1000008
74. Mäki-Tanila A, Hill WG (2014) Influence of gene interaction on complex trait variation with multilocus models. Genetics 198:355–367
75. Li J, Li X, Zhang S, Snyder M (2019) Gene-environment interaction in the era of precision medicine. Cell 177:38–44
76. Vivian-Griffiths T, Baker E, Schmidt KM, et al (2019) Predictive modeling of schizophrenia from genomic data: Comparison of polygenic risk score with kernel support vector machines approach. Am J Med Genet Part B Neuropsychiatr Genet 180:80–85
77. Bellot P, de Los Campos G, Pérez-Enciso M (2018) Can Deep Learning Improve Genomic Prediction of Complex Human Traits? Genetics 210:809–819. https://doi.org/10.1534/genetics.118.301298
78. Azodi CB, McCarren A, Roantree M, et al (2019) Benchmarking algorithms for genomic prediction of complex traits. bioRxiv 614479
79. Upton A, Trelles O, Cornejo-Garc\'ia JA, Perkins JR (2016) High-performance computing to detect epistasis in genome scale data sets. Brief Bioinform 17:368–379
80. Domingo J, Baeza-Centurion P, Lehner B (2019) The causes and consequences of genetic interactions (epistasis). Annu Rev Genomics Hum Genet 20:433–460
81. Young AI, Durbin R (2014) Estimation of epistatic variance components and heritability in founder populations and crosses. Genetics 198:1405–1416
82. Guindo-Mart\'inez M, Amela R, Bonàs-Guarch S, et al (2020) The impact of non-additive genetic associations on age-related complex diseases. bioRxiv
83. Brandes N, Linial N, Linial M (2021) Genetic association studies of alterations in protein function expose recessive effects on cancer predisposition. Sci Rep 11:14901. https://doi.org/10.1038/s41598-021-94252-y
84. Hunter DJ (2005) Gene--environment interactions in human diseases. Nat Rev Genet 6:287–298
85. McAllister K, Mechanic LE, Amos C, et al (2017) Current challenges and new opportunities for gene-environment interaction studies of complex diseases. Am J Epidemiol 186:753–761
86. Gauderman WJ, Mukherjee B, Aschard H, et al (2017) Update on the state of the science for analytical methods for gene-environment interactions. Am J Epidemiol 186:762–770
87. Kerin M, Marchini J (2020) Inferring Gene-by-Environment Interactions with a Bayesian Whole-Genome Regression Model. Am J Hum Genet 107:698–713
88. Kerin M, Marchini J (2020) A non-linear regression method for estimation of gene-environment heritability. Bioinformatics
89. Pirastu N, Cordioli M, Nandakumar P, et al (2021) Genetic analyses identify widespread sex-differential participation bias. Nat Genet 53:663–671
90. Weissbrod O, Flint J, Rosset S (2018) Estimating SNP-based heritability and genetic correlation in case-control studies directly and with summary statistics. Am J Hum Genet 103:89–99
91. Girirajan S, Campbell CD, Eichler EE (2011) Human copy number variation and complex genetic disease. Annu Rev Genet 45:203–226. https://doi.org/10.1146/annurev-genet-102209-163544
92. Weischenfeldt J, Symmons O, Spitz F, Korbel JO (2013) Phenotypic impact of genomic structural variation: insights from and for human disease. Nat Rev Genet 14:125–138
93. Payer LM, Burns KH (2019) Transposable elements in human genetic disease. Nat Rev Genet 20:760–772
94. Stancu MC, Van Roosmalen MJ, Renkens I, et al (2017) Mapping and phasing of structural variation in patient genomes using nanopore sequencing. Nat Commun 8:1–13
95. Rhoads A, Au KF (2015) PacBio Sequencing and Its Applications. Genomics Proteomics Bioinformatics 13:278–289. https://doi.org/https://doi.org/10.1016/j.gpb.2015.08.002
96. Choi Y, Chan AP, Kirkness E, et al (2018) Comparison of phasing strategies for whole human genomes. PLoS Genet 14:e1007308
97. Thorpe J, Osei-Owusu IA, Avigdor BE, et al (2020) Mosaicism in human health and disease. Annu Rev Genet 54:487–510
98. Nurk S, Koren S, Rhie A, et al (2021) The complete sequence of a human genome. bioRxiv
99. Voichek Y, Weigel D (2020) Identifying genetic

variants underlying phenotypic variation in plants without complete genomes. Nat Genet 52:534–540. https://doi.org/10.1038/s41588-020-0612-7

100. Li H, Feng X, Chu C (2020) The design and construction of reference pangenome graphs with minigraph. Genome Biol 21:1–19

101. DeRosse P, Karlsgodt KH (2015) Examining the psychosis continuum. Curr Behav Neurosci reports 2:80–89

102. Pies R (2007) How "objective" are psychiatric diagnoses?:(guess again). Psychiatry (Edgmont) 4:18

103. Graber ML (2013) The incidence of diagnostic error in medicine. BMJ Qual Saf 22:ii21--ii27

104. Cai N, Revez JA, Adams MJ, et al (2020) Minimal phenotyping yields genome-wide association signals of low specificity for major depression. Nat Genet 52:437–447. https://doi.org/10.1038/s41588-020-0594-5

105. Dahl A, Zaitlen N (2020) Genetic Influences on Disease Subtypes. Annu Rev Genomics Hum Genet 21:413–435

106. Dahl A, Cai N, Ko A, et al (2019) Reverse GWAS: Using genetics to identify and model phenotypic subtypes. PLoS Genet 15:e1008009